\documentclass{ws-procs9x6}

\def\be{\begin{equation}}
\def\ee{\end{equation}}
\def\beq{\begin{eqnarray}}
\def\eeq{\end{eqnarray}}

\begin{document}

\title{Nonextensive statistical effects on nuclear astrophysics and many-body problems}

\author{A. Lavagno and P. Quarati}

\address{Dipartimento di Fisica, Politecnico di Torino 
and INFN - Sezione di Torino e di Cagliari, Italy}

\maketitle

\abstracts{
Density and temperature conditions in many stellar core (like the solar core) 
imply the presence of nonideal plasma effects with 
memory and long-range interactions between particles. This aspect suggests the possibility that the 
stellar core could not be in a global thermodynamical equilibrium but satisfies the conditions of a 
metastable state with a stationary (nonextensive) power law distribution function among ions. The 
order of magnitude of the deviation from the standard Maxwell-Boltzmann distribution can be derived 
microscopically by considering the presence of random electrical microfields in the stellar plasma. 
We show that such a nonextensive statistical effect can be very relevant in many nuclear astrophysical 
problems. 
}

\section{Introduction}

The solar core is a neutral system of electron, protons, alpha particles and other heavier nuclei, usually assumed as 
an ideal plasma in thermodynamical equilibrium described by a Maxwellian ion velocity distribution. 
Because the nuclear rates of the most important reactions in stellar core are strongly affected by the high-energy 
tail of the ion velocity distribution, let us start by remanding the meaning of 
ideal and non-ideal plasma. 
A plasma is characterized by the value of the plasma parameter $\Gamma$
\be
\Gamma=\frac{\langle U\rangle_{\rm Coulomb}}{\langle T \rangle_{\rm thermal}} \; ,  
\ee
where $\langle U\rangle_{\rm Coulomb}$ is the mean Coulomb potential energy and 
$\langle T \rangle_{\rm thermal}$ is the mean kinetic thermal energy. Depending on the 
value of the plasma parameter we can distinguish three regimes: \\
\noindent
- $\Gamma\ll 1$ - Diluite weakly interacting gas, the Debye screening length 
$R_D$ is much greater than the average interparticle distance $r_0\approx n^{1/3}$, 
there is a large number of particles in the Debye sphere. \\
- $\Gamma\approx 0.1\div1$ - $R_D\approx r_0$, it is not possible to clearly separate 
individual and collective degree of freedom and the plasma is a 
weakly non-ideal plasma.\\
- $\Gamma\geq 1$ - High-density/low-temperature plasma, Coulomb interaction and 
quantum effects dominate and determine the structure of the system.

In the solar interior the plasma parameter $\Gamma_{\odot} \simeq 0.1$ and 
the solar core can be considered as a weakly nonideal plasma.
Similar behavior occurs in other astrophysical systems with $0.1<\Gamma<1$, 
among the others we quote brown dwarfs, the Jupiter core, stellar atmospheres.
Weakly nonideal conditions can influence how the stationary equilibrium can be reached 
within the plasma. 
In fact, in weakly nonideal astrophysical plasmas we have that the collision time is of the same 
order of magnitute of the mean time between collisions, 
therefore, several collisions are necessary before the particle loses memory
of the initial state; collisions between quasi-particles (ion plus screening cloud) are 
inelastic and {\bf long-range} interactions are present. 

In the next section we will see how the presence of memory and long-range forces can influence the thermodynamical stability 
and the stationary distribution function inside the stellar core.

\section{Metastable states of stellar electron-nuclear plasma}

We can distinguish two kind of thermodynamical equilibrium state \cite{sewell}:
\begin{itemize}
\item
global thermodynamical equilibrium: the free energy density is minimized globally
\item
local thermodynamical equilibrium: free energy density is minimized only in a 
restricted space, not globally. In this case the system is in a metastable state.
\end{itemize}

Metastable states are always characterized by {\bf long-range interactions and/or 
fluctuations} of intensive quantities (like inverse temperature 
$\beta$, density, chemical potential) and the stationary distribution function 
can be different from the Maxwellian one. In fact, in many-body long-range-interacting systems, 
it has been recently observed the emergence of long-standing {\it quasi stationary (metastable) states} 
characterized by non-Gaussian velocity distributions, before the Boltzmann-Gibbs equilibrium is attained 
\cite{latora,monte}.

Considering the corrections to an ideal gas due to identity of particles 
and to inter-nuclear interaction 
and the black-body radiation emitted, 
by minimizing the free energy density of the electron-nuclear plasma, 
we have obtained the following values \cite{meta}
\begin{displaymath}
n_{\ast }\simeq 2.74\cdot 10^{-14}\,\mathrm{fm}^{-3}\;\;,\;\;
\mathrm{k_{B}} T_{\ast }\simeq 5\cdot\,\mathrm{keV}\;\;\mathrm{and}\;\;R_{\ast }\approx 0.2R_{\odot }\;,
\end{displaymath}
with a typical stellar chemical composition $\bar{Z}=1.25$.
States with different values of $\mathrm{k_{B}}T$ (lower) and $n$ (higher) are 
{\bf metastable} states that can be featured by temperature fluctuations or density 
fluctuations, by quasi-particle models or by the presence of 
self-generated magnetic fields or random microfields distributions. 

The values obtained above are more than three times higher than the actual temperature 
of the solar interior and an electron density about half the actual value in the solar core. Therefore, 
the core of a star like the Sun can not exactly be considered in a global thermodynamical 
equilibrium state but can be better described as a metastable state and the stationary distribution function could be 
slightly different form the Maxwellian distribution.

In this context, it has been shown that when many-body long-range interactions are present, in many cases the system 
exhibits stationary metastable properties with power law distribution well described within 
the Tsallis nonextensive thermostatistics \cite{tsallis,borges,wei,ananos}.

\section{Microscopic interpretation: random electrical microfield}

In this section we want to investigate about a microscopic justification of a metastable power-law stationary 
distribution inside a stellar core. At this scope, let us start by observing that 
the time-spatial fluctuations in the particles positions produce 
specific fluctuations of the microscopic electric field 
(with energy density of the order of $10^{-16}$ MeV/fm$^3$) in a given point of the plasma.
These microfields   
have in general long-time and long-range 
correlations and can generate anomalous diffusion.
The presence of the electric microfield average energy 
density, $\langle E^2\rangle$, modifies the stationary solution 
of the Fokker-Planck equation and the ion equilibrium distribution can be 
written as 
\begin{equation}
f(v)=C \exp \left \{ -\int_0^v \frac{m v dv }
{kT (1+\frac{\langle E^2 \rangle}{E_c^2}}
\right \} \; ,
\label{distri}
\end{equation}
where $E_c=\nu \sqrt{3 x m kT/2 e^2}$.
In the solar core being $E$ not too larger than $E_c$, 
the distribution differs slightly from the Maxwellian.
Crucial quantity is the elastic collision cross section is the enforced elastic Coulomb cross section
$\sigma_0=2\pi(\alpha r_0)^2$  where $r_0$ is the inter-particle distance, 
$\alpha$ is related to the pair-correlation function $g(R,t)$. 
The stationary (metastable) distribution (\ref{distri}) for the solar interior can be written 
as a function of the kinetic energy $\epsilon_p$ 
\begin{equation}
f(\epsilon_p)=N \exp{\left[ - \frac{\epsilon_p}{kT}
- \delta  \left(\frac{\epsilon_p}{kT}\right)^2 
               \right]}  \, ,
\label{diclay2}
\end{equation}

where the deformation parameter $\delta=(1-q)/2$ can be written as \cite{plb2001}
\begin{equation}
\label{deltam}
\vert \delta \vert \approx 
\frac{\sigma_0^2}{3 \langle\sigma_d^2\rangle}=12 \, \alpha^4 
\, \Gamma^2 \ll 1 \, .
\end{equation}
A reasonable evaluation of $\alpha$ gives: $\alpha=0.55$, with 
$\Gamma \sim 0.1$ and we obtain $q=0.990$ ($\delta=0.005$). 
In the next section we will see as such a small deviation of the MB distribution can be very relevant in 
several nuclear astrophysical applications.

\section{Signals in astrophysical problems}

Let us illustrate few problems where we can find signals 
of the presence of deviations from the MB distribution. Their solutions can be achieved by means 
of modified (or generalized) rates calculated by means of deformed distributions. 
Among the others, we quote  
A) Solar neutrino fluxes;
B) Jupiter energy production;
C) Atomic radiative processes in electron nuclear plasmas;
D) Abundance of Lithium;
E) Temperature dependence of modified CNO nuclear reaction rates and resonant fusion reactions. 
For brevity we will discuss here the last point only. A detailed discussion of the other problems can be found in 
Ref.s \cite{plb2001,corra}.

\subsection{Temperature dependence of modified CNO nuclear reaction rates}

The temperature dependence of CNO cycles nuclear rates is strongly affected 
by the presence of nonextensive effects in Sun like stars evolving towards white dwarfs 
($10^7\div 10^8\,\mathrm{K}$).
Small deviations ($q=0.991$) from MB distribution strongly increase the rates and can explain the 
presence of heavier elements (Fe, Mg) in final composition of white dwarfs,  
consistently with recent limit of the fraction of energy the Sun produces via the CNO fusion 
cycle (neutrino constraints). 
We obtain that \cite{ferro}
i) the luminosity yield of the $pp$ chain is slightly affected by the deformed statistics, 
with respect to the luminosity yield of the CNO cycle;
ii) the nonextensive CNO correction ranges from $37\%$ to more than $53\%$;
iii) above $T\approx 2\cdot 10^7\,{\rm K}$, the luminosity is mainly due to the CNO cycle only, thus 
confirming that CNO cycle always plays a crucial role in the stellar evolution, 
when the star grows hotter toward the white dwarf stage. 
Our results are reported in Fig. 1 and Fig. 2. In Fig. 1, 
we plot the dimensionless luminosity over temperature, for the $pp$ chain and the CNO cycle. 
In Fig. 2, we report the dimensionless equilibrium concentrations of CNO nuclei over temperature.

\begin{figure}[ht]
\centerline{\epsfxsize=4.1in\epsfbox{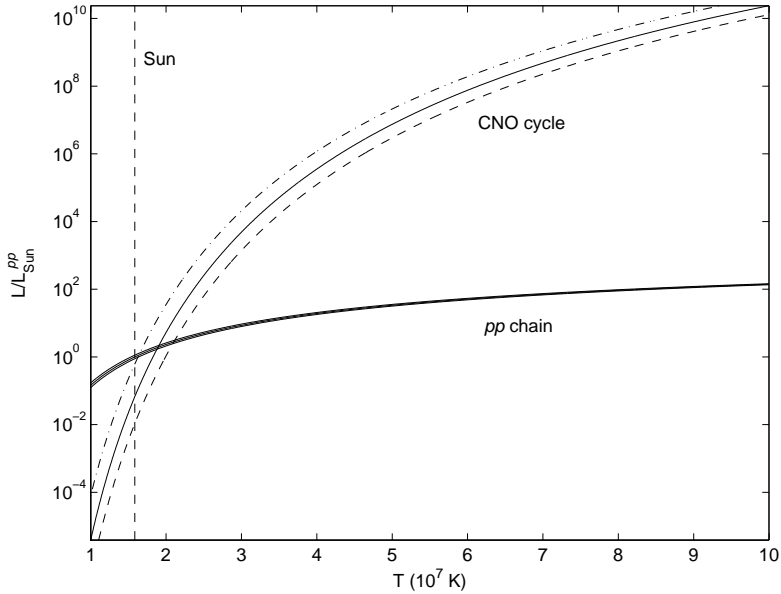}}   
\caption{Log-linear plot of dimensionless luminosity over temperature, for the $pp$ chain and the 
CNO cycle. Dashed line, $\delta=+0.0045$, $q=0.991$; dash-dotted line, $\delta=-0.0045$, $q=1.009$. The vertical line 
shows the Sun's temperature. All curves are normalized with respect to the $pp$ luminosity inside the Sun.}
\end{figure}

\begin{figure}[ht]
\centerline{\epsfxsize=4.1in\epsfbox{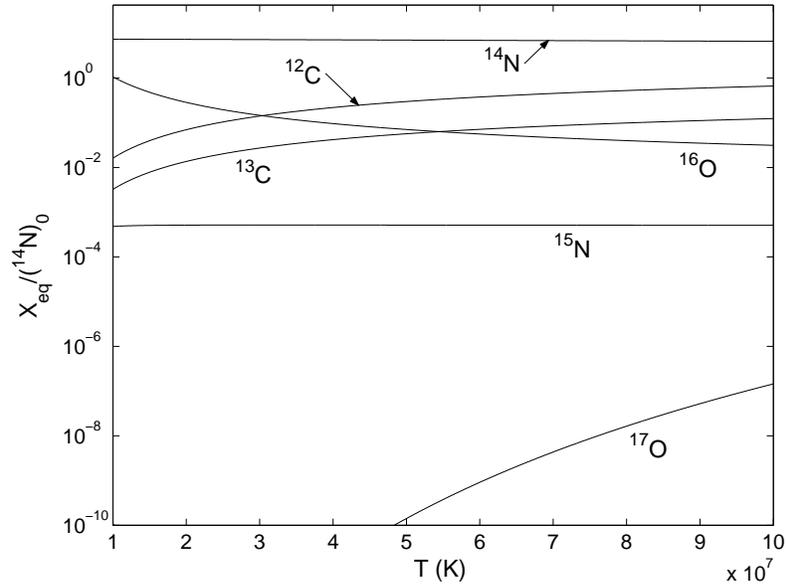}}   
\caption{Log-linear plot of dimensionless equilibrium concentrations of CNO nuclei over temperature. Classical 
statistics has been used. All curves are normalized with respect to the initial density $(^{14}N)_0$ inside the 
Sun.}
\end{figure}

\subsection{Resonant reaction rates in astrophysical plasma}

Cussons, Langanke and Liolios \cite{langa} proposed, on the basis of experimental measurements 
at energy $E\sim2.4\,\mathrm{MeV}$, that the resonant behavior of the stellar $^{12}\mathrm{C}+{^{12}\mathrm{C}}$ 
fusion cross section could continue down to the astrophysical energy range.

The reduction of the resonant rate due to resonant screening correction amounts to 11 orders of magnitude at the
resonant energy of $400\,\mathrm{keV}$, with important implications for hydrostatic
burning in carbon white dwarfs.

We have analytically derived two first-order formulae that can be used to express the non-extensive
reaction rate as a product of the classical reaction rate times a suitable corrective factor for both narrow and wide
resonances.

Concerning the fusion reactions between two medium-weighted nuclei, for example the $^{12}\mathrm{C}+{^{12}\mathrm{C}}$
reaction, our non-extensive factor, which can be formally defined as follows \cite{ferro2}
\begin{displaymath}
f_{NE}=1+\frac{15}{4}\delta-\left(\frac{E_R}{\mathrm{k_B} T}\right)^2\delta\; ,
\end{displaymath}
gives rise to further correction beside the screening and the potential resonant screening
\begin{equation}
F=f_{NE}\cdot f_S\cdot f_{RS}\; ,  \label{final corrective factor}
\end{equation}
where $f_S$ and $f_{RS}$ account for the Debye-H\"{u}ckel screening and the resonant screening effect respectively.

We have applied our results to a physical model describing a carbon white dwarf's plasma, with a temperature of
$T=8\cdot 10^8\,\mathrm{K}$ and a mass density of $\rho=2\cdot 10^9\,\mathrm{g/cm^3}$ (the plasma parameter is,
correspondingly, $\Gamma\simeq 5.6$). Furthermore, we have set a deformation parameter $|\delta|=10^{-3}$, regardless
of its sign, and we have kept the energy of the possible resonance, $E_R$, as a free parameter. 
In Fig. 3 we plot our estimation of the effective total factor $F$ as a function of the resonance energy $E_R$.

\begin{figure}[ht]
\centerline{\epsfxsize=4.1in\epsfbox{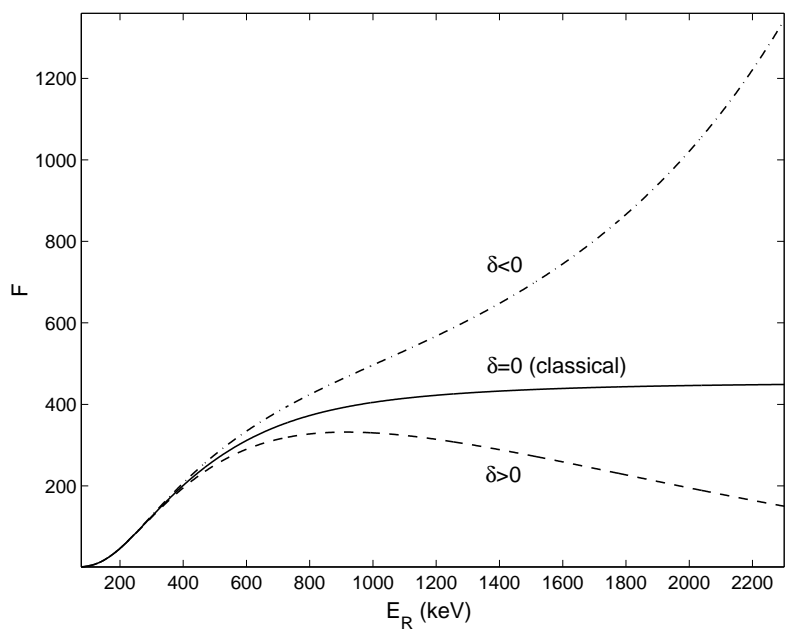}}   
\caption{Linear plot of the effective factor $F$, defined in Eq.(\ref{final corrective factor}), 
against the resonance energy $E_R$. The dash-dotted (upper) line refers to super-extensivity, 
the dashed (lower) line to sub-extensivity,
while the solid (middle) line describes the classical (MB) result.}
\end{figure}

All the plasma enhancements due to the presence of long-range many-body nuclear correlations and memory
effects
are in the direction of still more increasing the effective factor $F$ of nuclear rates of hydrostatic
burning and white dwarfs environment.

\section{Signals in high-energy nuclear collisions}

In this section we want briefly to remark as nonextensive statistical effects can be also very relevant 
also in the phenomenological interpretation of the high-energy nuclear collisions data. 
In fact, the quark-gluon plasma close to the critical temperature is a strongly interacting system. 
For such a system, the color-Coulomb coupling parameter of the QGP can be defined in analogy as 
\beq
\Gamma \approx  C \frac{g^2}{r_0\, T} >1 
\eeq
where $C=4/3$ or 3 is the Casimir invariant for the quarks or
gluons, respectively, and $\alpha_s=g^2/(4\pi)=0.2 \div 0.5$, 
$r_0\simeq n^{1/3}\simeq 0.5$ fm.  

Near the phase transition, the interaction range is much larger than the Debye screening length 
(small number of partons in the Debey sphere). In fact, 
$\lambda_D=1/\mu\le 0.2$ fm (using the non-perturbative
estimate: $\mu=6T$)v. 
The Coulomb radius for a thermal parton with energy $3T$
is given by $<r>=Cg^2/3T=1\div 6$ fm. Therefore one obtain $<r>/\lambda_D =5 \div 30$. 
Memory effects and long--range color interactions give rise to the presence 
of non--Markovian processes in the kinetic equation affecting the 
thermalization process toward equilibrium as well as the standard 
equilibrium distribution. 

A complete description of the applicability of nonextensive statistical effects to high-energy heavy ion 
collisions lies out the scope of this contribution. However, we want to outline that this aspect has been 
recently studied by us in connection to a phenomenological interpretation of the SPS data \cite{epjc,pla} 
and an analysis of the transverse pion momentum spectra and the net proton rapitity distribution 
measured at RHIC is under investigation.

\end{document}